\documentclass[preprint,aps,prl]{revtex4}
\usepackage{graphicx}
\def \ds {\displaystyle}
\def \bc {\bar c}
\def \bS {\bar S}
\def \bas {\bar s}
\def \te {\tilde \epsilon}

\def \bs {{\bf \sigma }}
\def \dsw {D_{{\rm sw}}}
\def \eke {E_{{\rm ke}}}
\def \wk {w_{{\bf k}}}
\def \TC {T_{\rm C}}
\def \JH {J_{\rm H}}
\def \vS {{\bf S}}
\def \vB {{\bf B}}
\def \vx {\hat {\bf x}}
\def \vy {\hat {\bf y}}
\def \vz {\hat {\bf z}}
\def \vQ {{\bf Q}}

\def \vs {{\bf s}}

\def \vk {{\bf k}}

\begin{document}

\title{Spin Dynamics of Double-Exchange Manganites with Magnetic Frustration}
\author{R.S. Fishman}
\affiliation{Condensed Matter Sciences Division, Oak Ridge National Laboratory, Oak Ridge, TN 37831-6032}

\begin{abstract}

This work examines the effects of magnetic frustration due to competing ferromagnetic and 
antiferromagnetic Heisenberg interactions on the spin dynamics of the double-exchange model.
When the local moments are non-colinear, a charge-density wave forms because the electrons
prefer to sit on lines of sites that are coupled ferromagnetically.  With increasing hopping
energy, the local spins become aligned and the average spin-wave stiffness increases.  Phase
separation is found only within a narrow range of hopping energies.  Results of this work 
are applied to the field-induced jump in the spin-wave stiffness observed in the manganite 
Pr$_{1-x}$Ca$_x$MnO$_3$ with $0.3 \le x \le 0.4$.

\end{abstract}

\maketitle

The persistence of antiferromagnetic (AFM) short-range order below the Curie temperarture 
$\TC $ of the manganites has been known for many years \cite{dag:01}.  Close to but below $\TC $,
metallic manganites like La$_{0.7}$Ca$_{0.3}$MnO$_3$ contain polaronic regions \cite{ada:00,koo:01} 
that are responsible for the coexistence of propagating and diffusive spin dynamics \cite{che:03}.  
In the remarkable compound Pr$_{1-x}$Ca$_x$MnO$_3$ with $0.3 \le x \le 0.4$, the low-temperature 
ferromagnetic (FM) insulating phase was originally believed to be a canted AFM (CAF) \cite{yos:95,oki:99} 
but probably contains regions with both FM and AFM short-range order \cite{deac:01,sim:02}.  
When an applied field $B$ exceeds about 3 T, the resistivity drops by several orders of magnitude \cite{yos:95}, 
the AFM regions shrink \cite{sim:02}, and the spin-wave (SW) stiffness $\dsw $ jumps by a factor 
of 3 \cite{fer:02}.  Despite the recognition that short-range AFM order plays a central role in the 
manganites, little is known theoretically about how long-wavelength SW's are affected by propagating through
both FM and AFM regions.  Because electron hopping is hampered by the misalignmnent of the local 
moments \cite{and:55}, AFM interactions may be expected to suppress the contribution of electron-mediated 
double-exchange (DE) to the SW dynamics \cite{deg:60}.  This paper examines the effects of 
AFM interactions and non-colinearity on the SW dynamics of electrons coupled to the local 
moments of a generalized Villain model \cite{vil:77,gab:89,sas:92}.  Our results strongly suggest that
the jump in the SW stiffness observed \cite{fer:02} in Pr$_{0.7}$Ca$_{0.3}$MnO$_3$ is produced by 
a sharp increase in the hopping energy at the critical field rather than by the alignment of the AFM regions.

As pictured in Fig.1(a), the local moments $\vS_i$ of the generalized Villain model are 
coupled by the FM interaction $J$ along the $\vy $ direction and by either the FM interaction $J$ 
or the AFM interaction $-\eta J$ along the $\vx $ direction.  A zero-temperature CAF phase is stabilized 
when $\eta $ exceeds $\eta_c$, which is $1/3$ when $\vB=B\vz = 0$ but increases as $B$ increases.  
Villain's original model \cite{vil:77} set $\eta =1$, which is the condition for full frustration.  
Due to the different environments of the $a$ and $b$ sites, the angle $\theta_b$ at the $b$ sites 
is always larger than $\theta_a$ at the $a$ sites, as shown in Fig.1(b).

\begin{figure}
\includegraphics *[scale=0.6]{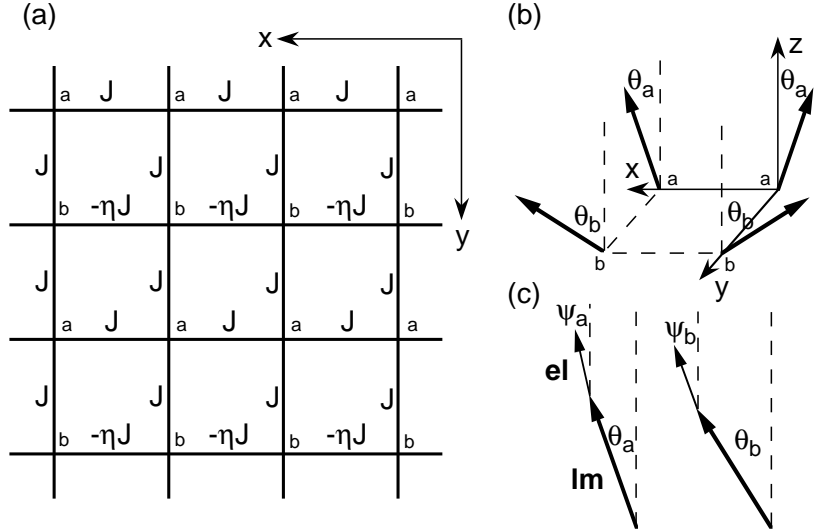}
\caption{
(a)  The generalized Villain model with Heisenberg couplings $J$ or $-\eta J$, (b) the
local moments in the $xz$ plane subtend angles $\theta_a$ and $\theta_b$ with the $z$ axis, 
and (c) the electron spins also lie in the $xz$ plane but subtend angles $\psi_a < \theta_a$
and $\psi_b < \theta_b $ with the $z$ axis.
}
\end{figure}  

Within our hybrid model, the Heisenberg interactions between the local moments 
are given by the generalized Villain model while electrons with density $p=1-x$ are FM coupled 
to the local moments by Hund's coupling $\JH$ and hop between neighboring sites with energy $t$.  
The DEV model (so called because it combines the DE and generalized Villain models) 
provides several advantages as a basis for understanding the effects of magnetic frustration
on the spin dynamics.  First, it is one of the simplest periodic models that is magnetically 
frustrated, which can be controlled through the parameter $\eta $.  In contrast to the case 
in a DE model with AFM interactions between all neighboring local moments \cite{kag:99,gol:00}, 
a homogeneous CAF phase is stable against phase separation \cite{inst} except in a very narrow 
region of parameter space.  Second, unlike a model with AFM exchange only, the DEV model supports 
FM order even when $t=0$ and $B=0$.  So it can be used to track the change in SW stiffness $\dsw $ 
as the electrons become mobile.  Third, because it contains both FM and AFM Heisenberg interactions, 
the DEV model can be used to study insulating manganites like Pr$_{0.66}$Ca$_{0.34}$MnO$_3$, where 
the AFM interactions arise from superexchange and the FM interactions from short-range orbital 
and polaronic order \cite{miz:00,pol}.  

For simplicity, our model is translationally symmetric with the FM and AFM Heisenberg couplings 
arranged periodically in two dimensions.  In the low-temperature phase of Pr$_{0.67}$Ca$_{0.33}$MnO$_3$, 
the FM interactions may be confined to two-dimensional sheets in a ``red cabbage'' structure \cite{sim:02}.  
But for wavelengths longer than the thickness $\sim 25 \AA $ of the FM sheets, the SW's will average over 
the FM and AFM regions.  So the DEV model will provide qualitiatively accurate predictions for the average 
SW stiffness $\dsw^{av}=(\dsw^x+\dsw^y)/2$, which is defined in the long-wavelength limit.
   
The Hamiltonian of the DEV model is 
\begin{equation}
\label{ham}
H=-t\sum_{\langle i,j \rangle }\sum_{\alpha }\Bigl( c^{\dagger }_{i\alpha }c^{\, }_{j\alpha }
+c^{\dagger }_{j\alpha }c^{\, }_{i\alpha } \Bigr) -2 \JH \sum_i \vs_i \cdot \vS_i
-\sum_{\langle i,j \rangle }J_{ij}\vS_i \cdot \vS_j -B\sum_i \bigl(S_{iz}+s_{iz}\bigr) ,
\end{equation}
where $c^{\dagger }_{i\alpha }$ and $c_{i\alpha }$ are the creation and
destruction operators for an electron with spin $\alpha $ at site $i$,
$\vs_i =(1/2) c^{\dagger }_{i\alpha } \bs_{\alpha \beta } c^{\, }_{i\beta }$
is the electronic spin, and $\vS_i$ is the spin of the local moment with magnitude $S$. 
The Heisenberg interactions $J_{ij}$ take the values $J$ or $-\eta J$, as described in
Fig.1(a).  This model will be solved at zero temperature to lowest order in $1/S$.  
To guarantee that the contributions to the SW frequencies from DE hopping $t$ and from the 
Heisenberg interactions $J_{ij}$ are of the same order in $1/S$, $t$ is considered to be of 
the same order in $1/S$ as $\JH S$, $JS^2$ and $BS$ (although their relative values can be 
quite different).  Thus, the dimensionless parameters of the DEV model are $t'=t/JS^2$, 
$\eta $, $B'=B/JS$, and $\JH /JS$.  To lowest order in $1/S$, the magnetic field 
$B$ only couples to the local moments.  While the theory developed below can be extended to treat
all values of the Hund's coupling, for simplicity we shall consider the limit of 
large $\JH S$ or in dimensionless terms, $\JH /JS \gg 1$ and $\JH S/t \gg 1$.

To solve this model, a Holstein-Primakoff expansion is first performed within the rotated 
reference frame of each spin:
$\bS_{iz} =S-a_i^{\dagger }a_i$, $\bS_{i+}=\sqrt{2S}a_i$, and $\bS_{i-}=\sqrt{2S}a_i^{\dagger }$.
In terms of electronic creation and destruction operators $\bc_{\vk ,\alpha }^{(r) \dagger }$
and $\bc_{\vk ,\alpha }^{(r)}$ in the rotated reference frame of the local moments, 
the zeroth-order term (in powers of $1/\sqrt{S}$) in the Hamiltonian can be written as
$H_0=E_h+H_b$ where to lowest order in $t/\JH S$,
\begin{equation}
\label{Eh}
E_h=\ds\frac{1}{2}NJS^2\biggl\{ -\cos 2\theta_a +\eta \cos 2\theta_b -2\cos (\theta_a-\theta_b)
-B'  \Bigl( \cos \theta_a +\cos \theta_b \Bigr) \biggr\},
\end{equation}
\begin{eqnarray}
\label{Hb}
H_b=&\ds\sum_{\vk ,\alpha } \biggl\{ \bc_{\vk \alpha }^{(a) \dagger }\bc_{\vk \alpha }^{(a)} \Bigl(
-\JH S \alpha -2t\cos k_x \cos \theta_a \Bigr)
+\bc_{\vk \alpha }^{(b) \dagger }\bc_{\vk \alpha }^{(b)} \Bigl(
-\JH S \alpha -2t\cos k_x \cos \theta_b \Bigr)
\nonumber \\ &
-\Bigl(\bc_{\vk \alpha }^{(a) \dagger }\bc_{\vk \alpha }^{(b)}   
+\bc_{\vk \alpha }^{(b) \dagger }\bc_{\vk \alpha }^{(a)}  \Bigr)
2t\cos k_y \cos \bigl( (\theta_a-\theta_b )/2 \bigr)\biggr\}.
\end{eqnarray}
Here, the lattice constant is set to 1 and $\alpha =\pm 1 $ corresponds to spin up or down in the
local reference frames.  

The electronic Hamiltonian $H_b$ is easily transformed
into the diagonal form $H_b=\sum_{\vk ,\alpha, r}\epsilon_{\vk \alpha }^{(r)}d_{\vk \alpha }^{(r)\dagger }
d_{\vk \alpha }^{(r)}$  by the rotations $\bc_{\vk \alpha }^{(a)}=u_{\vk }^{(a)}d_{\vk \alpha }^{(a)}
+u_{\vk }^{(b)}d_{\vk \alpha }^{(b)}$
and $\bc_{\vk \alpha }^{(b)}=u_{\vk }^{(b)}d_{\vk \alpha }^{(a)}-u_{\vk }^{(a)}d_{\vk \alpha }^{(b)}$
where $\epsilon_{\vk \alpha }^{(r)}=-\JH S \alpha +\te_{\vk }^{(r)}$, 
$u_{\vk }^{(a) 2}=1-u_{\vk }^{(b) 2} = \bigl(1 + (\cos\theta_a -\cos\theta_b)\cos k_x/\wk \bigr)/2$,
\begin{equation}
\te_{\vk }^{(r)}=-t\cos k_x \bigl(\cos \theta_a +\cos\theta_b \bigr) \mp t \wk ,
\end{equation}
\begin{equation}
\wk =\sqrt{\bigl(\cos \theta_a -\cos \theta_b\bigr)^2\cos^2 k_x +4\cos^2 \bigl( (\theta_a-\theta_b)/2\bigr) \cos^2 k_y}. 
\end{equation}
The $\mp $ signs refer to the $r=a$ and $b$ bands, respectively,
and the first Brillouin zone extends from $-\pi $ to $\pi $ in the $k_x$ direction but only
from $-\pi/2$ to $\pi /2 $ in the $k_y$ direction due to the reduced symmetry. 

In the limit of large $\JH S$, the zeroth-order energy $E_0=E_h+\langle H_b\rangle $ can readily be
minimized with respect to the angles $\theta_a$ and $\theta_b$.  When $t=B=0$, $\theta_b=3\theta_a$
for all $\eta $.  For a fixed $\eta $ and $B'$, the equilibrium angles decrease with increasing $t'$
and $\theta_b < 3\theta_a$.  The phase boundary between the CAF and FM phases satisfies the condition
\begin{equation}
\label{pb}
B'-2\eta +4 +3\eke /4JS^2-2\sqrt{ (1+\eta )^2+(1 +\eke /8JS^2)^2 }=0,
\end{equation}
where $\eke =-\bigl(\langle \te^{(a)}_{\vk }\rangle +\langle \te^{(b)}_{\vk }\rangle \bigr)/2 > 0$ is 
the average kinetic energy of the electrons in the FM phase.
For $p=0.66$, $\eta =3$, and $B=0$, the dependence of the equilibrium angles on $t'$ is plotted in 
Fig.2(a).  Also shown is the average spin $M=S(\cos\theta_a +\cos\theta_b)/2$ of the local moments.

\begin{figure}
\includegraphics *[scale=0.6]{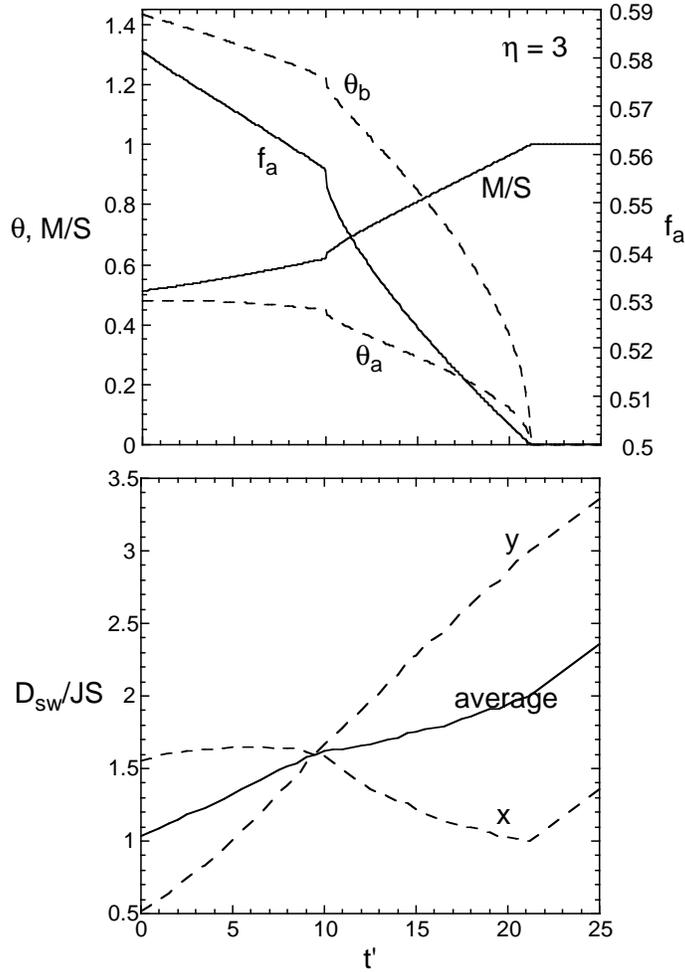}
\caption{
(a) The angles $\theta_a$ and $\theta_b$ of the local moments, the total local magnetization
$M/S$, the fraction $f_a$ of the electrons on the $a$ sites, and (b) the SW stiffnesses
for $p=0.66$, $\eta =3 $ and $B=0$ versus $t'$. 
}
\end{figure}

Surprisingly, the electronic occupation of the $a$ and $b$ sites are different with most of the 
electrons sitting on the $a$ sites.  The fraction $f_a$ of such electrons, also plotted in Fig.2(a), 
has a maximum of 0.58 as $t'\rightarrow 0$ and approaches 1/2 as $t'\rightarrow t'_c\approx 21.2 $.  
This behavior is easy to understand:  the largest angles between neighboring spins are along the $x$ 
axis between $b$ sites with angles differing by $2\theta_b $.  When an electron hops onto a $b$ site, 
it cannot easily hop to other $b$ sites and so quickly moves onto a neighboring $a$ site, where it can 
readily travel between other $a$ sites with angular difference $2\theta_a \ll 2\theta_b$.  Hence, 
the non-colinearity of the local moments quite naturally produces a charge-density wave (CDW) with 
a substantial amplitude.  A CDW with the same period as the one predicted here has in fact been 
observed in the insulating phase of Pr$_{0.7}$Ca$_{0.3}$MnO$_3$ \cite{oki:99,asa:02}.

Another surprise is that phase separation occurs within a narrow range of $t'$ around 10.0.  
Phase separation is easily seen in a plot of filling $p$ versus chemical potential $\mu $ 
as a discontinuity $\Delta p$ in $p(\mu )$.  It appears at fixed $p$ as jumps in the equilibrium angles $\theta_r$ 
and electron fraction $f_a$.  Like the Pomeranchuk instability \cite{hal:00} in the two-dimensional 
Hubbard model, the phase instability in the DEV model occurs close to a Van Hove filling and is marked 
by a change in Fermi surface (FS) topology from closed to open, as shown in the inset to Fig.3 where 
the FS is sketched for values of $t'$ on either side of the phase-separated range.  For $t'=10.2$, 
the extra electrons in the neck of the $a$ FS around $\vk =0$ are cancelled by the holes in the 
$b$ FS around $\vk =(\pi ,\pi/2)$.  However, the phase separation is extremely weak and for the 
parameters in Fig.2, $\Delta p \approx 0.003$. 

\begin{figure}
\includegraphics *[scale=0.6]{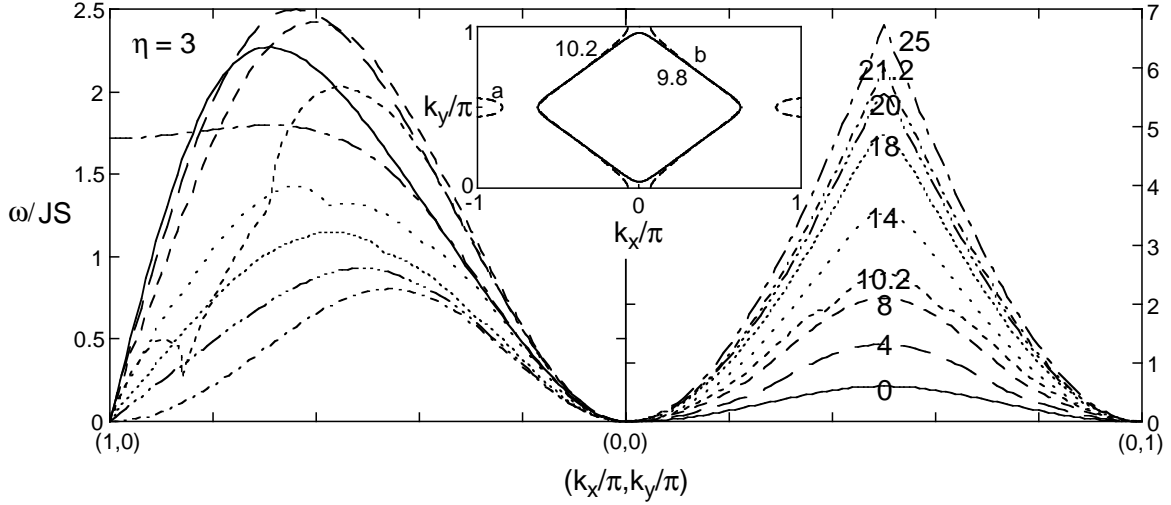}
\caption{
The SW frequencies for $p=0.66$, $\eta=3$, $B=0$, and various values of $t'$.  
Different energy scales are used on either side of $\vk =0$.
In the inset, we plot the FS for $t' =9.8$ (solid) and 10.2 (dashed), on either side of
a narrow region of phase separation.  
}
\end{figure}

As sketched in Fig.1(c), the equilibrium angles $\psi_r$ for the electrons are not equal to the angles 
$\theta_r$ of the local moments except when $\JH S /t =\infty $.  For finite $\JH S /t$, $\psi_r < \theta_r$ 
as the electrons try to align their spins as much as possible.  In the limit of large Hund's coupling, 
$\theta_r-\psi_r\propto t/\JH S$ and the electrons always exert a small torque on the local moments.  
Hence, the relationships given above for $\bc_{\vk \alpha }^{(r)}$ in terms of $d_{\vk \alpha }^{(s)}$ 
should contain admixtures of opposite-spin terms like $(t/\JH S)d_{\vk +\vQ, -\alpha }^{(s)}$, 
where $\vQ =(\pi ,0)$ is the AFM Bragg vector.  Since $\langle \bas_{ix} \rangle \propto 
\sin (\theta_i -\psi_i )$, these new terms produce a correction to the Hund's coupling 
$-2\JH \vS_i \cdot \vs_i $ that survives in the $\JH S \rightarrow \infty $ limit.  

After diagonalizing the band Hamiltonian $H_b$, the full Hamiltonian can be expanded as a power series 
in $1/\sqrt{S}$:  $H=H_0+H_1+H_2+\cdots $, where $H_1$ contains the torque terms and is linear in 
the boson operators $a_{\vk }^{(r)}$.  To eliminate this first-order term and to express the Hamiltonian 
in terms of the true SW operators for the total spin $\vS_{i, {\rm tot}}=\vS_i+\vs_i$, we perform the 
unitary transformation \cite{gol:00} $H'=e^{-U}He^U$ where $U$ is constructed to satisfy $[U,H_0]=H_1$.  
To lowest order in $1/S$, a trace over the Fermion degrees of freedom yields the modified 
second-order Hamiltonian $H_2'=H_2+[U,H_1]/2$ for the SW operators only: 
\begin{equation} 
H'_2=JS\sum_{\vk, r,s}\biggl\{ a_{\vk }^{(r)\dagger }a_{\vk }^{(s)}A_{\vk }^{(r,s)}
+\Bigl( a_{-\vk }^{(r)}a_{\vk }^{(s)}+a_{-\vk }^{(r)\dagger }a_{\vk }^{(s)\dagger }\Bigr)
B_{\vk }^{(r,s)} \biggr\}.
\end{equation}
It is then straightforward to diagonalize $H_2'$ to obtain the mode frequencies $\omega_{\vk }$.  
The torque terms discussed above are required to preserve rotational symmetry and the 
relations $\omega_{\vk =0}=B$ and $\omega_{\vQ }=0$ in the CAF phase. 

In the FM phase, the SW frequency is given by the analytic result 
\begin{eqnarray}
\omega_{\vk }= &B+JS\bigl( 3-\eta +(\eta -1)\cos k_x \bigr) +\eke (2-\cos k_x)/4S
\nonumber \\ &
-JS\sqrt{ (1+\eta )^2(1-\cos k_x)^2 +4(1+\eke /8JS^2)^2\cos^2 k_y }.
\end{eqnarray}
An AFM component develops when $\omega_{\vQ }=0$, which yields the same condition
for the phase boundary as Eq.(\ref{pb}).  For the FM phase, the SW stiffness obtained from 
the long-wavelength expansion $\omega_{\vk }\approx B+\dsw^x k_x^2 +\dsw^yk_y^2$ is simply the sum 
of the DE and Heisenberg contributions: $\dsw^x=\eke /8S+JS(1-\eta )/2$ and $\dsw^y=\eke /8S+JS$.  
For $J=0$, these results agree with the SW frequencies of the DE model first obtained by 
Furukawa \cite{fur:95}.

In the CAF phase, the SW frequency and stiffness must be solved numerically.
When $t=0$, our results agree with Saslow and Erwin \cite{sas:92} for the generalized Villain model.
Results for $\omega_{\vk }$ are plotted in Fig.3 for $p=0.66$, $\eta =3$, $B'=0$, and 
various values of $t'$.  Above the phase separation region around $t' \approx 10.0$ but
below $t'_c\approx 21.2$, $\omega_{\vk }$ develops kinks that correspond to transitions across the 
neck of the $a$ FS ($\vQ - \vk\approx 0.14 \pi \vx $ for $t'=10.2$) and the length of the 
$b$ FS ($\vQ - \vk \approx 0.31\pi \vx $ for $t'=10.2$).  

As plotted in Fig.2(b), the SW stiffness in the $\vx $ direction reaches a minimum at $t'_c$,
above which both $\dsw^x$ and $\dsw^y$ are linearly increasing functions of $t'$.
The stiffnesses in the $\vx $ and $\vy $ directions cross in the region of phase separation,
where the SW's are isotropic in the long-wavelength limit.  Notice that $\dsw^{av}$ 
increases by roughly a factor of 2 as $t'$ increases from zero to $t'_c$ and the system
transforms from a CAF with a CDW into a FM.
 
By contrast, the effect of a magnetic field is quite different.  After a sudden increase of the
SW stiffness for very small fields that occurs in any CAF \cite{fis:un}, 
there is a gradual increase in $\dsw^y$ and decrease in $\dsw^x$ 
as $B$ increases to $B_c$.  For $t'=3$ and $\eta =2$, $\dsw^{av}$ drops from
$1.2JS$ at $B=0$ to $0.54JS$ at $B_c$.  A field also very quickly eliminates the region
of phase separation.  

These results clearly indicate that the jump in SW stiffness observed \cite{fer:02}
in Pr$_{0.7}$Ca$_{0.3}$MnO$_3$ at a field of 3 T cannot be produced by simply aligning the AFM regions 
while keeping the bandwidth $\sim t$ fixed.  The identical resistivities in the metal-insulator transition
produced by either a magnetic field or x-rays \cite{kir:97} suggest a common mechanism:  
the excitation of charge carriers out of polaronic traps formed by the electron-lattice coupling.  
The doubling of $\dsw^{av}$ found in Fig.2(b) provides strong support for this 
scenario.  Since the integrated optical weight is proportional to $\eke \sim t$, 
the jump in the hopping energy at 3 T should be reflected in the optical conductivity. 
Measurements by Okimoto {\it et al.} \cite{oki:99} on Pr$_{0.6}$Ca$_{0.4}$MnO$_3$ do reveal a 
large increase in $\sigma (\omega )$ and a rapid drop in the CDW gaps near the critical
field.  If the percolation threshold for the FM regions \cite{mer:03} is exceeded when
the electrons delocalize, then the jump in the SW stiffness will coincide with the 
metal-insulator transition.  Otherwise the metal-insulator transition will occur 
at a slightly higher field. 

Because it requires two sublattices with filling $x=0.5$, local CE-type AFM ordering in the 
manganites Pr$_{0.6}$Ca$_{0.4}$MnO$_3$ \cite{oki:99,asa:02} and La$_{0.7}$Ca$_{0.3}$MnO$_3$ 
\cite{ada:00,koo:01} would be simplified if the polaronic regions were rich in holes and 
poor in electrons.  The DEV model provides a natural explanation for this behavior, since the 
electronic fraction on $b$ sites is substantially smaller than the fraction on $a$ sites as 
the electrons avoid regions with more pronounced AFM order.  

To conclude, we have studied the effect of AFM interactions on the magnetic order 
and SW dynamics of electrons interacting with the local moments of a generalized Villain 
model.  This model contains rich physics that provides insight into the SW dynamics of
any itinerant system with competing FM and AFM Heisenberg interactions.  
 
This research was sponsored by the U.S. Department of Energy under contract DE-AC05-00OR22725 with Oak Ridge 
National Laboratory, managed by UT-Battelle, LLC.  Conversations with Drs. A. Chernyshev, J. Fernandez-Baca, 
M. Katsnelson, N. Furukawa, W. Saslow, R. Wood, and A. Zheludev are happily acknowledged.


\begin{references}
\bibitem{inst}  The DEV model lacks the ``site-local continuous degeneracy''
that is responsible for the phase separation found in Ref.\cite{gol:00}.
\bibitem{pol} The FM coupling in the insulating phase of the manganites is associated with 
electronic hopping between neighboring Mn$^{3+}$ and Mn$^{4+}$ ions but is enhanced by the 
local polaronic distortions of the O atoms around some of the Mn$^{4+}$ ions \cite{miz:00}.  The 
FM Heisenberg coupling $J$ reflects the enhancement of the FM interaction beyond that produced by 
the uniform hopping of electrons on an undistorted lattice.
\end{references}
\end{document}